\documentclass{article}

\usepackage{arxiv}

\usepackage[utf8]{inputenc} 
\usepackage[T1]{fontenc}    
\usepackage{url}            
\usepackage{booktabs}       
\usepackage{amsfonts}       
\usepackage{nicefrac}       
\usepackage{microtype}      
\usepackage{lipsum}
\usepackage{authblk}
\usepackage[hidelinks]{hyperref}
\usepackage{graphicx}
\usepackage{amsmath,amssymb,bm}
\pdfoutput=1
\usepackage{tabularx}
\usepackage{changepage}

\usepackage{textcomp,marvosym}
\usepackage[right]{lineno}
\DisableLigatures[f]{encoding = *, family = * }
\usepackage[table]{xcolor}
\usepackage{array}

\newcolumntype{+}{!{\vrule width 2pt}}

\newlength\savedwidth

\makeatletter

\newcommand{\keywords}[1]{%
\par\noindent\textbf{Keywords: }#1
}

\title{A Herding-Based Model of Technological Transfer and Economic Convergence: Evidence from Central and Eastern Europe}

\author[2,3,*]{Vygintas Gontis\thanks{Corresponding author: \texttt{vygintas@gontis.eu}}}
\author[1]{Lesya Kolinets}

\affil[1]{Vilnius Gediminas Technical University (VILNIUS TECH), Lithuania.
\\ ORCID: \href{https://orcid.org/0000-0002-7005-0519}{0000-0002-7005-0519}
\\ \texttt{lesya.kolinets@vilniustech.lt}}

\affil[2]{Institute of Lithuanian Scientific Society, J. Basanavičiaus 6, Vilnius, Lithuania.
\\ ORCID: \href{https://orcid.org/0000-0002-1859-1318}{0000-0002-1859-1318}}

\affil[3]{Institute of Theoretical Physics and Astronomy, Vilnius University,
Saulėtekio al. 3, 10257 Vilnius, Lithuania}

\begin{document}
\maketitle

\begin{abstract}
The long-run convergence of developing economies toward advanced countries exhibits robust empirical regularities, yet the mechanisms underlying technological diffusion remain insufficiently specified in standard growth models. In this paper, we extend the neoclassical framework by introducing a micro-founded mechanism of technological transfer as a driver of total factor productivity. Rather than treating technological progress as exogenous or purely innovation-driven, we model productivity growth as a process of adopting existing technologies from the global frontier. The diffusion process is described using a herding-type interaction mechanism, in which agents transition from non-adopters to adopters under the combined influence of individual incentives and peer effects. This approach yields a tractable aggregate representation of TFP dynamics characterized by nonlinear convergence toward a moving technological frontier. We derive an explicit analytical solution and provide an interpretation of model parameters in terms of initial productivity, convergence limits, and diffusion speed. The model is evaluated using OECD productivity data for Central and Eastern European economies. 

\end{abstract}


\section{Introduction}

Economic growth and convergence across countries remain central topics in macroeconomics, with particular emphasis on the mechanisms enabling less developed economies to catch up with the technological frontier \cite{Solow1956,Swan1956,MankiwRomerWeil1992,Romer1990,AghionHowitt1992}. In the standard neoclassical framework, initiated by Solow, long-run growth is driven by exogenous technological progress embodied in total factor productivity (TFP). While this approach successfully explains steady-state properties, it leaves open the question of how technology diffuses across heterogeneous economies \cite{BarroSalaIMartin1997,NelsonPhelps1966,BenhabibSpiegel2005,CominHobijn2010,CominHobijn2004,AcemogluAghionZilibotti2006}.

Empirical evidence indicates that convergence is closely linked to the adoption of existing technologies rather than frontier innovation alone. Kremer (1993) formalized technological progress as proportional to the number of innovators, providing a scale-based mechanism for long-run growth. However, such models do not explicitly describe the process of technological transfer between countries operating at different levels of development.

An alternative perspective emphasizes diffusion-driven convergence, where productivity growth arises from the gradual adoption of advanced technologies. Despite its empirical relevance, this mechanism is typically introduced in reduced form. To address this limitation, recent studies employ agent-based approaches, where aggregate dynamics emerge from interactions among heterogeneous agents \cite{Aoki2002,Tesfatsion2006,GontisKononovicius2020}. In particular, Kirman's herding-type model  \cite{Kirman1993QJE}, offers a natural framework to describe the spread of technologies through local interactions and social influence \cite{Bass1969ManSci,Kononovicius2012IntSys}. Models have been successfully applied in economic and financial contexts, including prior work on agent-based and stochastic modeling of markets \cite{Gontis2014PlosOne}.

In this paper, we propose an extension of the neoclassical growth model in which TFP evolves through technological transfer from more advanced economies \cite{AcemogluAghionZilibotti2006}. The adoption process is modeled using a herding mechanism, where agents transition from non-adopters to adopters under the combined influence of individual incentives and interaction effects. This micro-founded approach yields a tractable macroscopic representation of productivity dynamics characterized by convergence toward a moving technological frontier.

We apply the model to OECD productivity data for Central and Eastern European (CEE) economies and estimate the key parameters governing the speed of technological transfer. CEE demonstrates rapid income catch-up episodes observed in the accession process to the EU and participation in global economic exchange, joining other organizations of economic cooperation \cite{Holobiuc2020EFAJ,Alemu2024AO}. Nevertheless, the transformation of CEE countries remains heterogeneous, and the comparison of their economic success has exceptional value for the further development of economic growth theory. The proposed framework captures observed convergence patterns while remaining parsimonious, thereby contributing to the integration of macroeconomic growth theory and agent-based modeling.

In Section \ref{sec:Model} we present the agent-based model of technology transfer, in Section \ref{sec:EmpiricalData} we analyze empirical data on CEE productivity, and in Section \ref{sec:Discussion} we discuss and conclude the results.

\section{Agent-based model for the technology transfer \label{sec:Model}}

Let us consider the neoclassical model of economic growth with two factors of productivity: labor $L$ and capital $K$
\begin{equation}
Y=AL^{1-\alpha}K^{\alpha},
\label{eq:Growth}
\end{equation}
where $A$ denotes the total factor productivity quantifying the technological advancement of the economy, we will refer to it as technological productivity. In the most simple global macro-historical approach, Kremer proposed \cite{Kremer1993QJEcon} a very simple model of permanent technological productivity growth generated by the total number of innovators $N(t)$
\begin{eqnarray}
\dot{A} & = & \gamma N(t) A(t),\label{eq:Kremer}\\
g_A(t) & = & \frac{\dot{A}}{A} = \gamma N(t), \label{eq:ARate}
\end{eqnarray}
where $g_A(t)$ is the growth rate and $\dot{A}$ denotes the time derivative of $A$. Equations assume the constant contribution $\gamma$ of any single innovator.

Countries in the World usually have very different levels of development reflected in the various values of technological productivity $A$. We seek to construct a dynamic model of $A$ for a single country based on technology transfer from more advanced countries. Such a time scale for development has to be much shorter than that for the development of fundamentally new technologies. Thus, we suggest replacing Eq. \eqref{eq:ARate} with the following one
\begin{equation}
g_A(t) =  \frac{\dot{A}}{A} = \gamma (N-X(t)), \label{eq:ARate2}
\end{equation}
where $X(t)$ denotes the number of innovators who have already accepted the most advanced technologies and implemented them. In this setup, we assume the total number of innovators $N$ in the country is constant and introduce a continuous variable $x(t)=X(t)/N$ for large $N$. For the modeling of $x(t)$, we use an extensive version of the herding model \cite{Kononovicius2012IntSys} previously used in the modeling of financial markets \cite{Gontis2014PlosOne} in its non-extensive form.
The transition rates $\pi^{\pm}(x)$, transition probabilities per unit time, in the non-extensive form, are used as follows
\begin{eqnarray}
\pi^{+}(x) &=& (1-x) \left(\frac{\sigma_1}{N} + h x \right) ,\label{eq:pip}\\
\pi^{-}(x) &=& x \left(\frac{\sigma_2}{N} + h [1-x] \right) .\label{eq:pim}
\end{eqnarray}
Here the $+$ denotes the transition $X->X+1$, and $-$ the transition $X->X-1$. 
Kirman's one-step discrete transition probabilities have been expressed
in the following form \cite{Kirman1993QJE},
\begin{eqnarray}
p( X \rightarrow X+1) &=& (N-X) \left(\sigma_1 + h X \right) \Delta t ,\label{eq:pp}\\
p( X \rightarrow X-1) &=& X \left(\sigma_2 + h [N-X] \right) \Delta t .\label{eq:pm}
\end{eqnarray}
The relation between the discrete transition probabilities,
(\ref{eq:pp}) and (\ref{eq:pm}), and continuous transition rates, (\ref{eq:pip})
and (\ref{eq:pim}), should be evident:
\begin{equation}
p ( X \rightarrow X \pm 1) = N^2 \pi^{\pm}(x) \Delta t,
\label{eq:disccont}
\end{equation}
where the transition time $\Delta t$ has to be small to ensure the opinion change of only one agent.

\begin{figure*}[!t]
\centering
\includegraphics[width=2.5in]{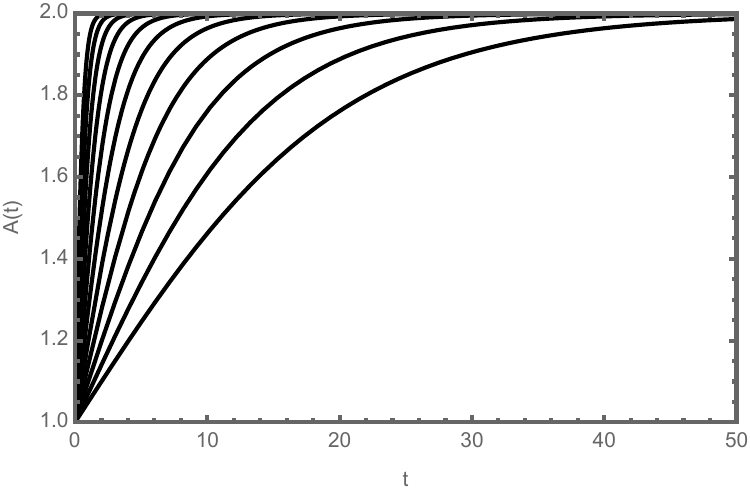}
\caption{Visualization of technological productivity model \eqref{eq:A-Solution-Final} with $A_0=1$, $A_m=2$, and $h=0.05 \times 2^{\frac{i}{2}}$, where $i=\{0,1,2....10\}$.}
\label{fig1}
\end{figure*}

For modeling technology transfer, one can use a simpler, extensive version (agents interact locally) of the herding model with one-directional transitions: agents accept new technologies but do not reject them. In such a case, we write the transition rates as follows
\begin{eqnarray}
\pi^{+}(x) &=& (1-x) \left(\frac{\sigma}{N} + \frac{h}{N} x \right) ,\label{eq:piptech}\\
\pi^{-}(x) &=& 0 .\label{eq:pimtech}
\end{eqnarray}
Using these rates, one can write the Master equation, or its approximation as a Fokker-Planck equation. Let us skip these steps (derivation you can find in \cite{Kononovicius2012IntSys}), writing the final result: the differential equation describing the dynamics of the continuous variable $x(t)$ 
\begin{eqnarray}
\mathrm{d} x &=& (1-x) (\sigma + h x) \mathrm{d} t, \label{eq:Bass}\\
\mathrm{d} s &=&  -s(\sigma+h (1-s))\mathrm{d} t , \label{eq:Bass-s}
\end{eqnarray}
where we introduced the variable change $s=1-x$ in the second equation. The explicit form for the solution of Eq. \eqref{eq:Bass-s} with initial condition $s(0)=1$ is as follows
\begin{equation}
s(t) = \frac{h+\sigma}{h+\sigma e^{(\sigma+h)t}} ,
\label{eq:Bass-Solution}
\end{equation}
The substitution of this solution into Eq. \eqref{eq:ARate} gives us the differential equation for $A(t)$
\begin{equation}
\frac{\mathrm{d} A}{A} = \gamma s(t) \mathrm{d} t = \gamma \frac{h+\sigma}{h+\sigma e^{(\sigma+h)t}} \mathrm{d} t,
\label{eq:A-Equation}
\end{equation}
which has the solution with one more constant of integration $\beta$
\begin{equation}
A(t) = \beta \frac{(h+\sigma) e^{(\sigma+h)t}}{1+\frac{\sigma}{h} e^{(\sigma+h)t}}.
\label{eq:A-Solution}
\end{equation}
We suggest to simplify the solution introducing limit values of $A(t)$: a) $A(0)=A_0$, and b) $A(t=\infty)=A_m$. Simplification gives us the final version for the model of technological productivity $A(t)$ with a clear meaning of its parameters: $A_0$, the initial value of technological productivity;  $A_m$, the maximum value of technological productivity equal to the productivity of the most advanced country; $h$, the time scale parameter defining the speed of convergence. Thus, we propose the following version of the technological productivity model 
\begin{equation}
A(t) = \frac{A_m A_0}{A_m e^{(-\frac{A_m h t}{A_m-A_0})}+ A_0 (1-e^{(-\frac{A_m h t}{A_m-A_0})})}.
\label{eq:A-Solution-Final}
\end{equation}
In Fig. \ref{fig1} we present visualization of proposed technological productivity model with $A_0=1$ and $A_m=2$ and the set of $h=0.05 \times 2^{\frac{i}{2}}$ with $i=\{0,1,2....10\}$.

Eq. \eqref{eq:A-Solution-Final} gives us a version for the model of technological productivity $A(t)$. The version of the model Eq. \eqref{eq:A-Solution-Final} has one shortcoming for empirical analyses, as $A_m$ is constant, in clear contradiction with empirical data: advanced economies experience technological development and GDP growth. Thus, we have to assume at least the simplest version of exponential growth $A_m=A_m^0 e^{(\gamma_m t)}$, solving a more sophisticated version of $A(t)$ logistic growth with moving carrying capacity
\begin{equation}
\mathrm{d} A = \gamma A(t) \left( 1-\frac{A(t)}{A_m^0 e^{(\gamma_m t)}}\right)  \mathrm{d} t,
\label{eq:A-Equation-2}
\end{equation}
Fortunately, this ordinary differential equation has an explicit solution as well
\begin{equation}
A(t) = \frac{A_m^0 e^{(\gamma t)} (\gamma-\gamma_m)}{\gamma (e^{(\gamma-\gamma_m) t}-1) + \frac{A_m^0}{A_0}
(\gamma-\gamma_m)}.
\label{eq:A-Solution-Final-2}
\end{equation}
We use this model of technological transfer for the further empirical analysis of economic convergence in Central and Eastern Europe with the advanced economies of Western Europe \cite{Kolinets2025-qb}.

\section{Empirical data analysis  \label{sec:EmpiricalData}}
\begin{figure*}[!t]
\centering
\includegraphics[width=5in]{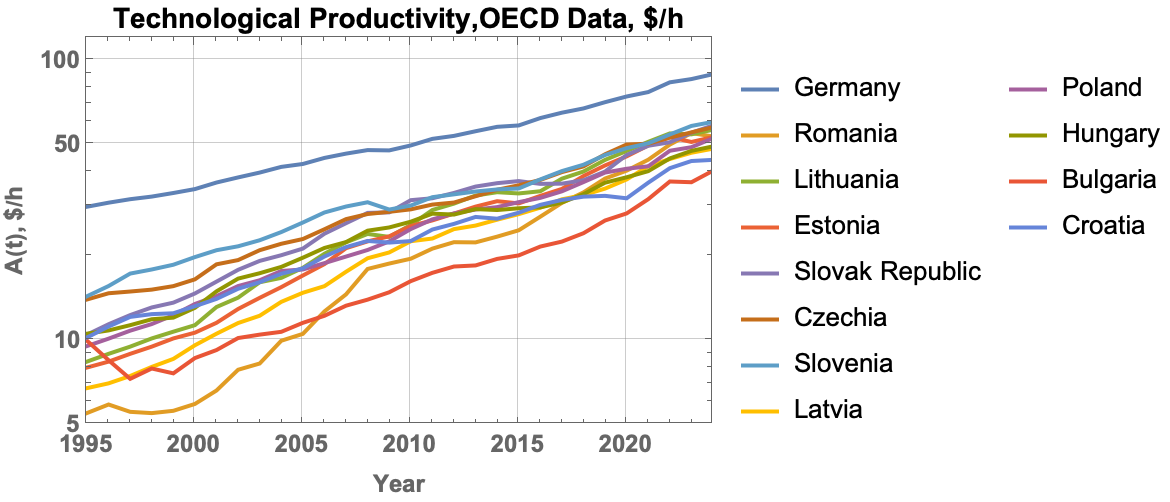}
\caption{Visualization of OECD data for the Total factor productivity of Germany and CEE countries. US dollars per hour worked, PPP converted, Current prices.}
\label{fig2}
\end{figure*}

For the empirical data analysis we use
the OECD Productivity Database providing users with the most comprehensive and the latest productivity estimates.  The productivity annual database contains data on total factor productivity measured using hours worked and the components of capital and labor inputs, US dollars per hour worked, PPP converted, Current prices. Further information for all data sets and the methodology may be found in the OECD tech report \cite{oecd2024productivity}. For the analysis we have chosen Germany as a reference country for catching up economies of Central and Eastern Europe (CEE): Bulgaria, Croatia, Czechia, Estonia, Hungary, Latvia, Lithuania, Poland, Romania, Slovakia, Slovenia. This choice is related to our previous research \cite{Kolinets2025-qb} seeking to extend the interpretation of the main economic growth factors. In Fig. \ref{fig2} we plot selected OECD data in log scale to demonstrate observed convergence of CEE economies. 

\begin{figure*}[!t]
\centering
\includegraphics[width=3.3in]{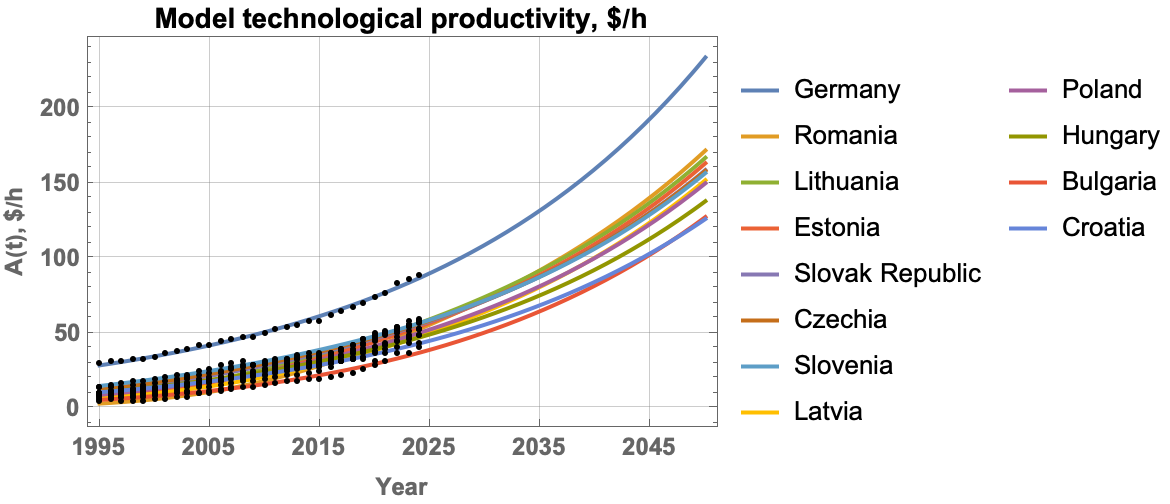}
\includegraphics[width=2.1in]{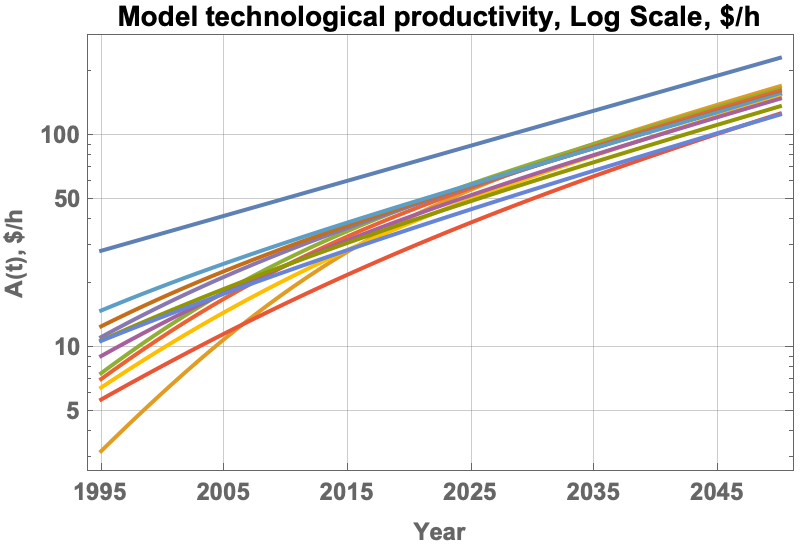}
\caption{Visualization of the Eq. \eqref{eq:A-Solution-Final-2} fitting to the OECD data procedure. Left figure shows the data points and fitting curves in linear scale, and the right figure demonstrates fitting curves in log scale.}
\label{fig3}
\end{figure*}

Our task is to find how the proposed model fits the data and projects future dynamics. We fit the exponential growth curve $A_m=A_m^0 e^{(\gamma_m t)}$ to the technological productivity data of Germany first. Then, using defined parameters $A_m^0=28.7205$ and $\gamma_m=0.0381261$, we fit Eq. \eqref{eq:A-Solution-Final-2} to the data of each CEE country and define their parameters $A_0$ and $\gamma$. We use least squares as fitting method — specifically, minimize the sum of squared residuals between the model and the data. We employ a Levenberg–Marquardt algorithm, a blend of gradient descent and Gauss–Newton methods. In Fig. \ref{fig3} we give the visualization of the fitting procedure.
In Table \ref{table1} we present fitting parameters, $A_0$, and $\gamma$, and projected values of technological productivity $A_{2030}$, and $A_{2050}$ in 2030 and 2050 respectively for the CEE countries catching up Germany. The main measure of proposed model is the parameter $\gamma$ defining the pace of long term growth. One can see that the order of countries in the table coincide with the magnitudes of $\gamma$ and the technological productivity $A_{2050}$. Nevertheless, the simplicity of the model giving us only one parameter responsible for the long term economic growth can be considered as a shortcoming, which eliminates the dependence on many other macroeconomic factors.

\begin{table}[h] 
\centering
\caption{Fitting parameters $A_0$ and $\gamma$ for the Eq. \eqref{eq:A-Solution-Final-2} adjusting it to the empirical data of the CEE countries catching up Germany, and projected values of technological productivity $A_{2030}$, and $A_{2050}$ for the years 2030 and 2050. Standard errors of parameters we denote as StdErr.}
\begin{tabularx}{\textwidth}{m{0.7cm}>{\centering\arraybackslash}X>{\centering\arraybackslash}X>{\centering\arraybackslash}X>{\centering\arraybackslash}X>{\centering\arraybackslash}X>{\centering\arraybackslash}X>{\centering\arraybackslash}X>{\centering\arraybackslash}X>{\centering\arraybackslash}X}
\midrule
\multicolumn{1}{>{\centering\arraybackslash}X}{{\textbf{Country}}} & \multicolumn{1}{>{\centering\arraybackslash}X}{{$\bm{A_0}$}} & \multicolumn{1}{>{\centering\arraybackslash}X}{{StdErr}} & \multicolumn{1}{>{\centering\arraybackslash}X}{{$\bm{\gamma}$}} & \multicolumn{1}{>{\centering\arraybackslash}X}{{StdErr}} & \multicolumn{1}{>{\centering\arraybackslash}X}{{$\bm{A_{2030}}$}} & \multicolumn{1}{>{\centering\arraybackslash}X}{{$\bm{A_{2050}}$}} \\ \midrule
{\textbf{Romania}}  & $3.25365$ & $0.314761$ & $0.148995$ & $	0.00464697$ & $72.7955$  & $171.836$   \\
{\textbf{Lithuania}}   & $7.58554$ & $0.336468$ & $0.136149$ & $0.00267282$ & $74.3754$ & $167.028$  \\ 
{\textbf{Estonia}}    & $7.11872$ &	$0.353558$ & $0.130107$ & $0.00286485$ & $71.7945$ & $163.379$   \\
{\textbf{Slovakia}}  & $11.239$ & $0.631024$ & $0.120902$ & $0.00390888$ & $71.7115$ & $158.83$  \\ 
{\textbf{Czechia}} & $12.6282$ & $0.410532$ & $0.119671$ & $0.00241689$ & $72.0422$ & $158.344$    \\ 
{\textbf{Slovenia}}  & $14.9978$ & $0.650397$ & $0.116691$ & $0.00358649$ & $72.1028$ & $156.821$  \\
{\textbf{Latvia}}   & $6.49365$ & $0.200544$ & $0.115141$ & $0.00167061$ & $64.438$ & $152.087$   \\ 
{\textbf{Poland}}  & $9.14255$ &	$0.265604$ & $0.110311$ & $0.00177447$ & $65.8209$ & $150.017$  \\
{\textbf{Hungary}}  & $10.8251$ & $0.483776$ & $0.0965172$ & $0.00288759$ & $61.1909$ & $138.089$  \\ 
{\textbf{Bulgaria}} & $5.69932$ & $0.382148$ & $0.0942085$ & $0.00344184$ & $50.6926$ & $127.524$  \\  
{\textbf{Croatia}}   & $10.7837$ & $0.357209$ & $0.0863183$ & $0.00211604$ & $55.861$ & $126.203$   \\ 

\bottomrule
\end{tabularx}
\label{table1}
\end{table}

Results presented in the Table \ref{table1} are dependent  on the choice of Germany as the reference country. Other similar models of long term projection choose United States (US) as reference for the long term catch-up process. Thus, in the Table \ref{table2} we present the same fitting parameters and growth projection for the CEE countries, when the US is chosen as a catching-up reference with empirical parameters $A_m^0=39.908$ and $\gamma_m=0.0354031$. The change of reference country makes some changes in the country ranking according the main parameter $\gamma$ value, as one can see in the Table \ref{table2}.

\begin{table}[h] 
\centering
\caption{Fitting parameters $A_0$ and $\gamma$ for the Eq. \eqref{eq:A-Solution-Final-2} adjusting it to the empirical data of the CEE countries catching up United States, and projected values of technological productivity $A_{2030}$, and $A_{2050}$ for the years 2030 and 2050. Standard errors of parameters we denote as StdErr.}
\begin{tabularx}{\textwidth}{m{0.7cm}>{\centering\arraybackslash}X>{\centering\arraybackslash}X>{\centering\arraybackslash}X>{\centering\arraybackslash}X>{\centering\arraybackslash}X>{\centering\arraybackslash}X>{\centering\arraybackslash}X>{\centering\arraybackslash}X>{\centering\arraybackslash}X}
\midrule
\multicolumn{1}{>{\centering\arraybackslash}X}{{\textbf{Country}}} & \multicolumn{1}{>{\centering\arraybackslash}X}{{$\bm{A_0}$}} & \multicolumn{1}{>{\centering\arraybackslash}X}{{StdErr}} & \multicolumn{1}{>{\centering\arraybackslash}X}{{$\bm{\gamma}$}} & \multicolumn{1}{>{\centering\arraybackslash}X}{{StdErr}} & \multicolumn{1}{>{\centering\arraybackslash}X}{{$\bm{A_{2030}}$}} & \multicolumn{1}{>{\centering\arraybackslash}X}{{$\bm{A_{2050}}$}} \\ \midrule
{\textbf{Romania}}  & $3.77933$ & $0.276553$ & $0.125784$ & $	0.00334579$ & $77.4339$  & $192.199$   \\
{\textbf{Lithuania}}   & $8.37812$ & $0.30658$ & $0.105743$ & $0.00196187$ & $77.3495$ & $177.999$  \\ 
{\textbf{Estonia}}    & $7.78863$ &	$0.329619$ & $0.103664$ & $0.00221673$ & $74.5066$ & $174.483$   \\
{\textbf{Latvia}}   & $6.90862$ & $0.195649$ & $0.0963823$ & $0.00143334$ & $66.335$ & $161.941$   \\ 
{\textbf{Slovakia}}  & $11.8308$ & $0.585946$ & $0.0907588$ & $0.00295312$ & $72.9275$ & $162.422$  \\ 
{\textbf{Poland}}  & $9.50355$ &	$0.226165$ & $0.089073$ & $0.00131183$ & $67.2848$ & $156.059$  \\
{\textbf{Czechia}} & $12.9852$ & $0.334397$ & $0.0888178$ & $0.00159386$ & $73.2775$ & $160.98$    \\ 
{\textbf{Bulgaria}} & $5.80356$ & $0.34917$ & $0.0841821$ & $0.00294871$ & $51.8$ & $134.609$  \\ 
{\textbf{Slovenia}}  & $15.1175$ & $0.55593$ & $0.0839232$ & $0.00358649$ & $72.6617$ & $156.019$  \\
{\textbf{Hungary}}  & $11.039$ & $0.452043$ & $0.0779928$ & $0.00237168$ & $61.702$ & $139.667$  \\ 
{\textbf{Croatia}}   & $10.8943$ & $0.336535$ & $0.0714743$ & $0.00178679$ & $56.0708$ & $126.407$   \\ 

\bottomrule
\end{tabularx}
\label{table2}
\end{table}
 
\section{Discussion and conclusions  \label{sec:Discussion}}
In this paper, we proposed a novel extension of the neoclassical growth framework in which total factor productivity evolves through technological transfer from more advanced economies. By introducing a micro-founded adoption mechanism based on herding interactions, we provided a tractable representation of how technology diffusion at the agent level translates into aggregate productivity dynamics. The resulting model generates a nonlinear convergence path toward a moving technological frontier and offers a transparent interpretation of key parameters governing the speed of economic growth.

The empirical application to OECD productivity data for Central and Eastern European economies demonstrates that the proposed specification captures essential features of observed convergence. As shown in the fitted trajectories, Figure \ref{fig3} and parameter estimates, Tables \ref{table1}, and \ref{table2}, differences in long-run growth across countries can be summarized by a small set of parameters, in particular the diffusion-driven growth rate $\gamma$ . This parsimonious structure is advantageous for empirical implementation and allows for consistent cross-country comparison of convergence dynamics.

At the same time, the simplicity of the model implies certain limitations. First, the framework abstracts from institutional, structural, and policy-related factors that are known to influence technological adoption. Second, the assumption of a homogeneous population of agents and a single diffusion channel may oversimplify the complexity of real-world innovation systems. Third, the specification of the technological frontier, whether represented by Germany or the United States, affects parameter estimates and projected dynamics, highlighting the sensitivity of results to the choice of benchmark economy.

Despite these limitations, the model provides a useful bridge between neoclassical growth theory and agent-based approaches. It offers a micro-founded interpretation of technological diffusion that complements existing reduced-form convergence models and connects naturally with empirical observations. In particular, the explicit link between interaction dynamics and aggregate productivity growth contributes to the broader literature at the intersection of macroeconomics and complex systems.

Future research may extend the framework in several directions. First, we would like to introduce the financial component through the rate of diffusion, $\gamma$, into the dynamics of TFP, as our previous work reveals the importance of private debt in the growth of CEE \cite{Kolinets2025-qb}. Incorporating heterogeneous agents, multiple sectors, or network structures could provide a more realistic description of technology diffusion. Finally, integrating additional macroeconomic variables and institutional indicators may improve empirical performance and enhance the explanatory power of the model.  

In conclusion, the proposed technological transfer framework offers a parsimonious and analytically tractable approach to modeling economic convergence. By embedding a diffusion-based mechanism within the neoclassical structure, it provides new insights into the role of technology adoption in shaping long-run growth dynamics.

\section{Abbreviations}
The following abbreviations are used in this manuscript:\\

\noindent 
\begin{tabular}{@{}ll}

CEE & Central and Eastern Europe\\
GDP & Gross Domestic Product\\
OECD & Organisation for Economic Co-operation and Development\\
PPP & Purchasing Power Parity\\
TFP & Total Factor Productivity\\

\end{tabular}

\bibliography{EconomicGrowth}

\end{document}